\documentclass[aps,prb,twocolumn,showpacs,amsmath,amssymb,preprintnumbers,10pt,article,natbib]{revtex4-2}
\usepackage{latexsym}
\usepackage{color}
\usepackage{graphicx}
\usepackage{soul}
\usepackage{makecell}
\usepackage{xcolor}
\usepackage{hyperref}
\hypersetup{
    colorlinks=true,
    linkcolor=black,
    urlcolor=blue,
    citecolor=blue
}

\begin {document}

\title{Atypical ferrimagnetism in  Ni$_4$Nb$_2$O$_9$}
\author{Jhuma Sannigrahi$^1$}
\email{jhuma@iitgoa.ac.in}
\author{Roumita Roy$^1$} 
\author{Richard Waite$^2$}
\author{Anupam Banerjee$^3$}
\author{Mohamad Numan$^4$}
\author{Manh Duc Le$^2$}
\author{D. T. Adroja$^2$}
\author{Sudipta Kanungo$^1$}
\email{sudipta@iitgoa.ac.in}
\author{Subham Majumdar$^4$} 
\email{sspsm2@iacs.res.in} 

\affiliation{$^1$School of Physical Sciences, Indian Institute of Technology Goa, Goa-403401, India}
\affiliation{$^2$ISIS Neutron and Muon Source, Science and Technology Facilities Council, Rutherford Appleton Laboratory, Didcot OX11 0QX, United Kingdom}
\affiliation{$^3$Department of Physics, Ram Ratan Singh College, Patliputra University, Mokama, Patna- 803302, India}
\affiliation{$^4$School of Physical Sciences, Indian Association for the Cultivation of Science, 2A \& B Raja S. C. Mullick Road, Jadavpur, Kolkata 700032, India}

\date{\today}

\begin{abstract}
Ferrimagnetism typically emerges from chemically distinct magnetic ions or the same element at two inequivalent crystallographic sites, rendering unequal moments. In contrast, Ni$_4$Nb$_2$O$_9$ has been recently discovered to show a different mechanism, where identical Ni$^{2+}$ ions with the same ligand coordination develop unequal magnetic moments purely due to differences in local environments. Here, we investigate the microscopic origin of this emergent mechanism through a synergy of powder neutron diffraction, inelastic neutron scattering, and first-principle-based calculations. We demonstrate that the Ni$_A$ and Ni$_B$ sublattices, while sharing the same nominal valence, differ in their magnetic dimensionality: Ni$_A$ forms quasi-one-dimensional chains with enhanced $p$–$d$ hybridization and a reduced magnetic moment, whereas Ni$_B$ retains a nearly two-dimensional geometry and a full $S =$ 1 moment. Our results underscore the pivotal role of spin dimensionality and local structural distortions in stabilizing ferrimagnetism in systems with electronically equivalent magnetic ions.

\end{abstract}
 
\maketitle

At the atomic level, ferrimagnetism arises when oppositely polarized spins have unequal populations or amplitudes, resulting in net magnetization ($M$)~\cite{FI1}. At the crystalline solid level, the imbalance of $M$ from oppositely aligned spins typically originates from differences in magnetic elements, valence states, or the crystallographic environment~\cite{FI2, FI3} possess the magnetic moments. But what happens when the same magnetic element with the same valence and spin state occupies inequivalent crystallographic sites with antiparallel spin alignment? This situation typically occurs in a collinear antiferromagnetic state with zero net magnetization. If this situation occurs with non-zero net magnetization, it results in an atypical ferrimagnetic (FI) state. This state breaks time-reversal symmetry and, to the best of our knowledge, is not widely recognized. Variations in the local structural environment can induce moment differences at two sites despite their identical valence states; nevertheless, experimentally confirming this phenomenon and understanding its underlying mechanisms remains a challenge. In this letter, we reveal the microscopic nature of the magnetic ground state in Ni$_4$Nb$_2$O$_9$, a material that exemplifies such a counterintuitive mechanism of ferrimagnetism in general.

\par
Ni$_4$Nb$_2$O$_9$ (NNO) crystallizes in an orthorhombic structure with the non-centrosymmetric space group \textit{Pbcn} (No.60). The Ni ions occupy two inequivalent crystallographic sites, designated Ni$_A$ and Ni$_B$, both of which have the same nominal valence state of 2+ ($S =$ 1). The Ni$_A$O$_6$ octahedra form quasi-one-dimensional zigzag chains within distinct planes via edge sharing, with inter-chain coupling mediated by corner-sharing [Fig.~\ref{fig:PND}(a)]. In contrast, edge-sharing Ni$_B$O$_6$ octahedra form nearly uniform two-dimensional honeycomb layers, with Ni$_B$ atoms remaining approximately coplanar. These sublattices are interconnected through face-sharing interactions, where each Ni$_B$ honeycomb layer couples to three Ni$_A$ chains above and below. The long-range magnetic order in NNO is observed below $\sim$ 76~K ($T_{FI}$), consistently reported as an FI ground state in earlier studies~\cite{NBO1, NBO2, NBO3, NBO4, NBO5,NBO6, fita}. Furthermore, a magnetic compensation point ($T_{cmp}$) appears around 33 K, below which negative magnetization is observed, which is a characteristic of magnetization reversal at the two magnetic sublattices in ferrimagnets. Thus, the ferrimagnetism in NNO arises from the imbalance in ordered magnetic moments at the two Ni sites, both of which are in octahedral coordination with six surrounding oxygen ions, and each unit cell contains an equal number of Ni$^{2+}$ ions.

\par
NNO was first reported to exhibit an FI ground state by Ehrenberg \textit{et al.} ~\cite{NBO2} in 1995 using powder neutron diffraction (PND), suggesting two ferromagnetically aligned Ni sublattices coupled antiferromagnetically. However, the precise magnetic moments at the Ni$_A$ and Ni$_B$ sites remained unresolved, and no further studies emerged for nearly 25 years. Recent investigations~\cite{NBO1, NBO3, NBO4, NBO5} since 2020 have reconfirmed the ground state of FI with moments aligned along the $b$ axis, validated the valence states of Ni$^{2+}$ and Nb$^{5+}$ (excluding mixed-valence scenarios) and demonstrated magnetization reversal below $T_{cmp}$. However, the microscopic origin of ferrimagnetism in NNO, especially the reason for the distinct magnetic moments at the two Ni sites, has remained unclear. Thota \textit{et al.}~\cite{NBO1} first attempted to explain this phenomenon by modelling the system using N\'eel's FI framework, extracting the molecular field exchange constants Ni$_{AA}$, Ni$_{BB}$, and Ni$_{AB}$, which represent interactions between Ni$^{2+}$ ions within and between the A and B sublattices. Their model reproduced temperature-dependent magnetization data by assuming significantly different Land\'e - $g$-factors (2.47 and 2.10) at the two sites, resulting in differing magnetic moments despite identical Ni$^{2+}$ valence states. In their study on Zn-doped NNO systems, Fiori \textit{et al.}~\cite{fiori} reported different Landé g-factor values for the two inequivalent Ni sites. However, such a large difference in Land\'e - $g$-factor requires substantially different spin-orbit coupling (SOC) strengths, which is unusual for Ni$^{2+}$ ions given the quenched orbital angular momentum in the $d^8$ configuration. Indeed, our first-principles calculations indicate that SOC has a negligible influence on the electronic structure, motivating us to explore alternative mechanism. Here, we propose that ferrimagnetism in NNO emerges from subtle but critical differences in local structural distortions at the two Ni sites, leading to distinct electronic structures and varying degrees of $p$–$d$ hybridization. Furthermore, the magnetic dimensionalities of the Ni$_A$ and Ni$_B$ sublattices play a crucial role in stabilizing the ground state of FI. To uncover the microscopic origin of this atypical FI behavior, we have conducted a comprehensive investigation that combines high-resolution PND, inelastic neutron scattering (INS), and density functional theory (DFT) calculations. This letter provides fundamental insights into the intricate interplay among structure, electronic correlations, and magnetism in this unique system.

\begin{figure}[t]
\centering
\includegraphics[width = 8cm]{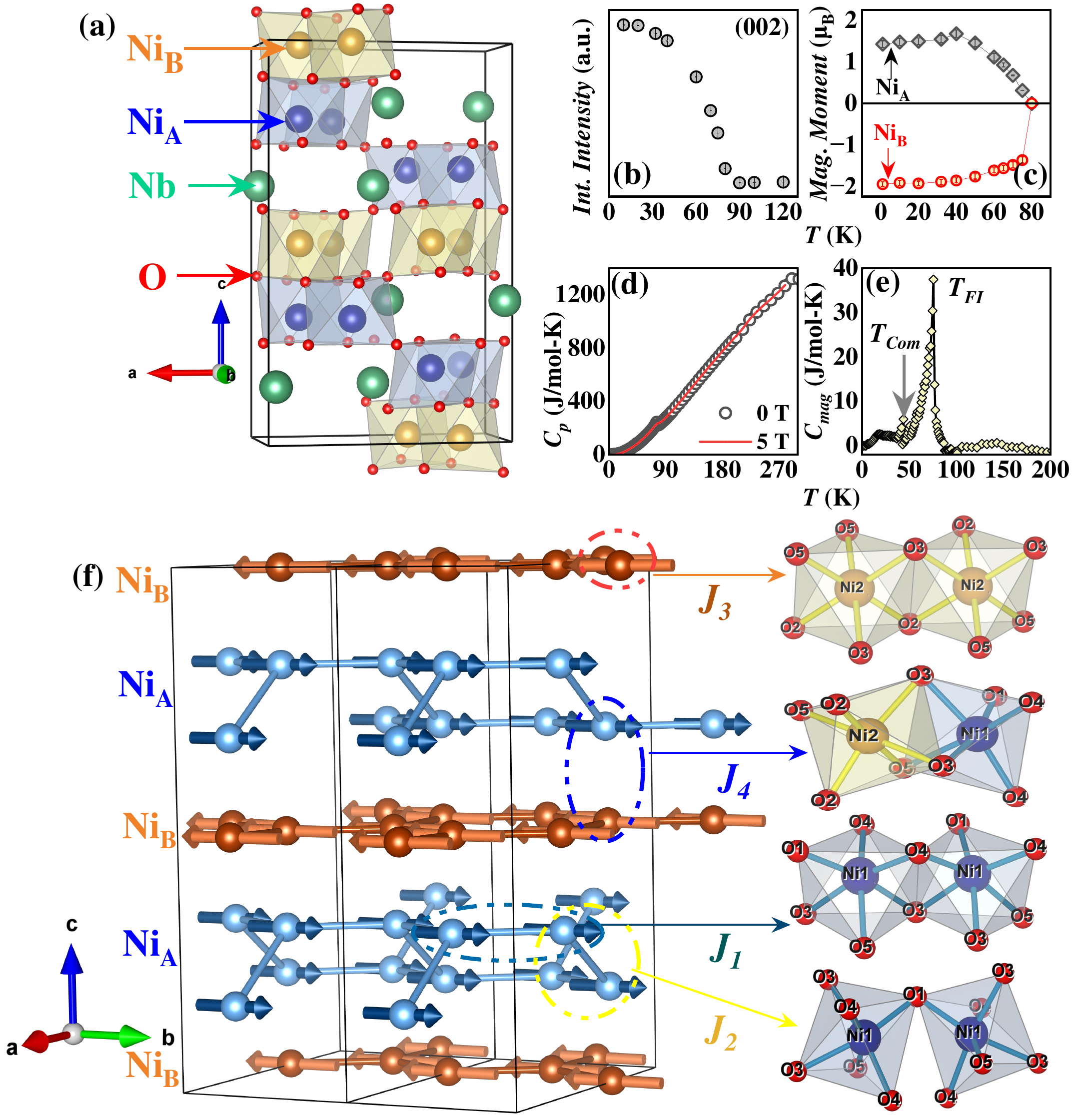}
\caption {(color online)(a) Perspective view of the crystal structure of NNO. (b) Temperature dependence of the integrated intensity of the (002) reflection with a significant magnetic contribution. (c) Temperature dependence of the ordered magnetic moments at the Ni$_A$ and Ni$_B$ sites. (d) Temperature dependence of heat capacity measured at applied magnetic fields of 0 and 5~T. (e) Magnetic contribution to the heat capacity. Further details are provided in the Supplementary Material~\cite{SM}. (f) Perspective view of the magnetic structure of NNO, along with the illustrated exchange interaction pathways.}  
\label{fig:PND}
\end{figure}  

PND study on NNO reveals that all magnetic reflections appear on top of the nuclear Bragg peaks and can be indexed using the propagation vector $\mathbf{k} =$ (0 0 0), consistent with previous reports~\cite{NBO3}. The temperature-dependent integrated intensity of the (002) peak, which exhibits a strong magnetic contribution, confirms the onset of long-range magnetic order below $T_{FI}$. Magnetic symmetry analysis indicates FM alignment within Ni$_A$ and Ni$_B$ sublattices, with ordered moments oriented along the $b$-axis [Fig.~\ref{fig:PND}(f)]. At 1.5 K, the refined magnetic moments are 1.429(48) $\mu_B$ for Ni$_A$ and 1.953(61) $\mu_B$ for Ni$_B$, coupled AFM along the $c$-axis. No significant canting from the $b$-axis is observed within the experimental uncertainty. Despite the quasi-one-dimensional and two-dimensional character of the Ni$_A$ and Ni$_B$ sublattices, respectively, the system exhibits robust three-dimensional long-range magnetic order below $T_{FI}$, with no evidence of short-range correlations in the diffraction data. Although the Ni$_B$ moment approaches the theoretical value for Ni$^{2+}$ (2 $\mu_B$), the Ni$_A$ moment remains somewhat reduced. The unequal and oppositely aligned moment values on the two Ni sites give rise to FI, as also reflected in the bulk magnetization data. The magnetic moment at the Ni$_A$ site increases below $T_{FI}$, shows an anomaly near 40 K, and then saturates at lower temperatures. In contrast, the Ni$_B$ moment steadily increases and saturates without anomalies. The spontaneous magnetization ($M_S$), obtained from isothermal magnetization data, exhibits a clear temperature dependence- reaching a maximum near 65K (just below $T_{FI}$) and vanishing around $T_{cmp}$~\cite{SM}. Heat capacity measurement [Fig.~\ref{fig:PND} (d-e)] further supports this behavior, showing two distinct $\lambda-$like anomalies at $T_{FI}$ and $T_{cmp}$, most prominent in the magnetic contribution of the heat capacity data. 

\begin{figure*}[t]
\centering
\includegraphics[width = 17 cm]{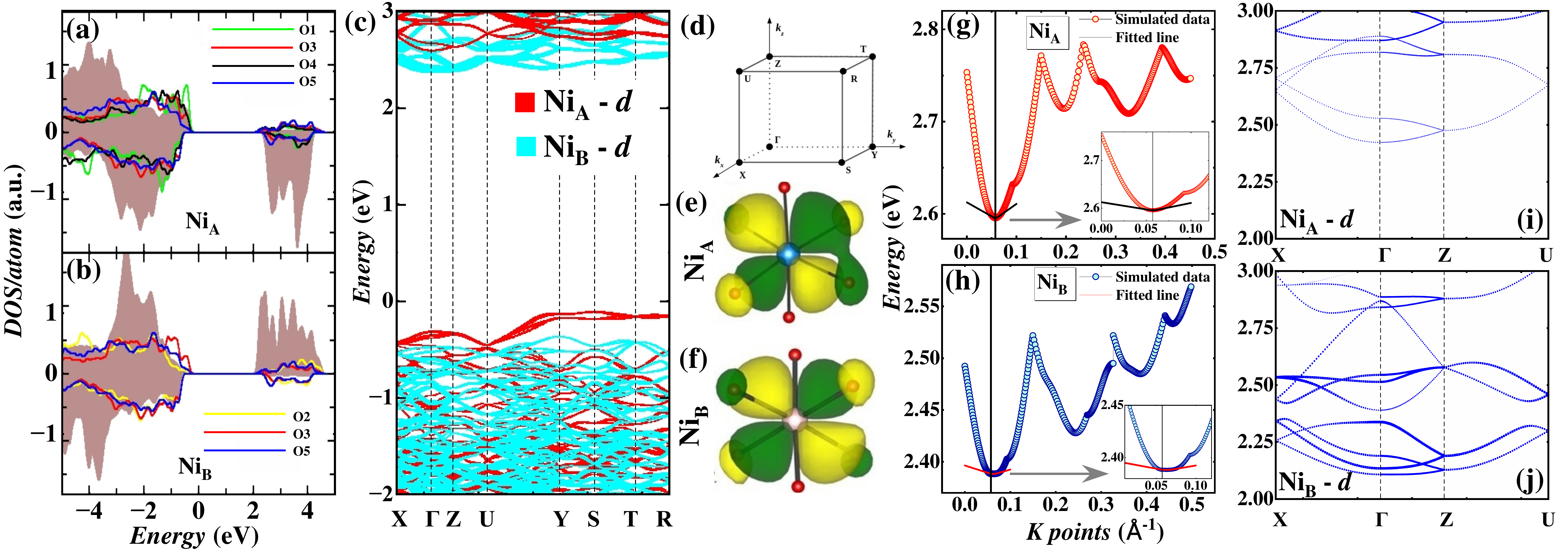}
\caption {(colour online) (a) and (b) show the site projected GGA+$U$ calculated DOS of the Ni$_A$ and Ni$_B$, respectively. The shaded DOS represents the $Ni-d$ states, while the colored lines represent the $O-2p$ states. (c) Shows the corresponding band structure of Ni$_A$ and Ni$_B$ in the specified BZ, which is shown in (d). (e) and (f) represents the effective Wannier function plot for Ni$_A$ and Ni$_B$, respectively. (g) and (h) shows the section of the conduction band minima at the $\Gamma$ point and the corresponding quadratic fitting of $E \propto k^2$ for Ni$_A$ and Ni$_B$, respectively. (i) and (j) represent the orbital projection of Ni-$d$ character weightage at the conduction band minima.} 
\label{fig:DFT}
\end{figure*}    

\par
We analyze the effective valency and local environments of Ni$_A$ and Ni$_B$ using bond valence sum (BVS) calculations from PND data~\cite{SM}. The valency of Ni$_B$ remains close 
to 2+ throughout the temperature range, whereas Ni$_A$ shows significant deviations that peak near $T_{FI}$ and gradually approach 2+ as the temperature approaches $T_{cmp}$, where the net magnetization vanishes. This trend closely mirrors the temperature dependence of magnetization, suggesting a strong correlation between charge redistribution and magnetic ordering. The oxygen ions surrounding the two Ni sites exhibit distinct behavior: O3, O4, and O5—coordinated to Ni$_B$—maintain valencies near –2, while O1, O2, O4, and O5—bonded to Ni$_A$—show marked deviations. This indicates enhanced charge transfer to the surrounding oxygens from Ni$_A$ compared to Ni$_B$. This charge transfer reduces the magnetic moment at Ni$_A$ compared to Ni$_B$. Furthermore, below $T_{FI}$, Ni$_A$-O bond lengths vary non-uniformly, with some bonds shortening (e.g., Ni$_A$-O1, Ni$_A$-O3) and others lengthening (e.g., Ni$_A$-O4). In contrast, Ni$_B$ -O bonds show uniform changes, reflecting a less distorted environment (see FIG. S3 in supplementary material).


To corroborate our experimental findings, we performed first-principles density functional theory (DFT) calculations on NNO. The results confirm that the FI configuration is energetically favored over the FM state by $\sim$74.5 meV/f.u., in agreement with the PND data. The calculated density of states (DOS) and band structure [Fig.~\ref{fig:DFT} (a-b)] reveal an insulating ground state with a band gap of $\sim$2 eV. Due to distorted octahedral coordination, the Ni-$d$ orbitals split into lower-energy $t_{2g}$ and higher-energy $e_g$ levels. The $t_{2g}$ states are fully occupied, while the $e_g$ states are partially filled in the majority spin channel and empty in the minority, yielding calculated spin moments of 1.71 and -1.72~$\mu_B$ for Ni$_A$ and Ni$_B$, respectively. Although both Ni sites possess a $d^8$ configuration with $S =$ 1, the DOS and band structure reveal notable sublattice differences. Ni$_A$ exhibits a narrower $d$ bandwidth, resulting in flat bands along the $Y$-$S$-$T$-$R$ path [Fig.~\ref{fig:DFT}(c)], while Ni$_B$ shows broader and more dispersive bands. Near the Fermi level, the valence band maximum is dominated by Ni$_A$ states, while the conduction band minimum is primarily derived from Ni$_B$ states. To understand magnetic exchange interactions, we calculated~\cite{Helberg1999, Mazurenko2006,Xiang2011,Majumder2015,Jhuma2019,Roy,fiori} four primary pathways [Fig.~\ref{fig:PND}(f)]: intra-chain edge-sharing Ni$_A$-Ni$_A$ ($J_1$), inter-chain corner-sharing Ni$_A$-Ni$_A$ ($J_2$), intra-plane edge-sharing Ni$_B$-Ni$_B$ ($J_3$), and inter-sublattice face-sharing Ni$_A$-Ni$_B$ ($J_4$). Within the Ni$_A$ sublattice, the strong FM intra-chain coupling $J_1 =$ –4.16 meV dominates over the weaker AFM inter-chain $J_2 =$ 0.78 meV, supporting quasi-one-dimensional correlations. In contrast, the Ni$_B$ sublattice hosts only a single FM interaction ($J_3 =$ –2.35 meV), forming a nearly planar honeycomb network. Crucially, the strong inter-sublattice AFM coupling $J_4 =$ 18.65 meV drives the overall FI ground state, resulting in a net moment due to the imbalance in the ordered magnetic moment between the two Ni sublattices. Despite the expected quenching of orbital degrees of freedom in the $d^8$ configuration of Ni$^{2+}$ and weak spin-orbit coupling (SOC), calculations predict a small orbital moment of 0.15~$\mu_B$/site for both Ni$_A$ and Ni$_B$, with negligible impact on the electronic structure. Further insight into the electronic structure was obtained through the low-energy tight-binding model in the maximally localized Wannier basis [Fig.~\ref{fig:DFT} (e) and (f)], which reveals a significant overlap between the orbitals of Ni 3$d$ and O 2$p$  at the Ni$_A$ site, indicative of strong $p$–$d$ hybridization. In contrast, such an overlap is negligible at the Ni$_B$ site. As a consequence, notable induced magnetic moments were observed on the oxygen atoms surrounding Ni$_A$: O1 (0.066 $\sim$ $\mu_B$), O4 (0.058~$\mu_B$), and O2 (0.049~$\mu_B$). This enhanced hybridization suggests an increase in charge transfer from Ni$_A$ to surrounding oxygen, leading to a reduced magnetic moment at the Ni$_A$ site compared to Ni$_B$, where a weaker hybridization preserves a higher magnetic moment - in agreement with the BVS results. 
\par
To further explore the electronic environment, we calculated the effective masses of the electrons ($m^*$) for Ni$_A$ and Ni$_B$ by fitting the bottom of the conduction band~\cite{effmass1,effmass2,effmass3} using a hyperbolic function [Fig.~\ref{fig:DFT}(g) and (h)]. The effective mass was determined from the curvature at the conduction band minimum $E_{\mathrm{cond}}(k_{\mathrm{CBM}})$ using the relation, $\frac{1}{m^*} = \frac{1}{\hbar^2} \frac{\partial^2 E_{\mathrm{cond}}}{\partial k^2}$. The flatter nature of the conduction band bottom at Ni$_B$ [insets of Fig.~\ref{fig:DFT}(g) and (h)] corresponds to a larger effective mass. Specifically, we find $m^*_{\mathrm{Ni_A}} = 0.1509\, m_e$ and $m^*_{\mathrm{Ni_B}} = 0.8824\, m_e$, where $m_e$ is the free electron mass. This reduced effective mass at Ni$_A$ is due to the weaker $d$-character and stronger hybridization, as also seen in the projected band structures [Fig.~\ref{fig:DFT}(i) and (j)]~\cite{effmass2}. The quasi-one-dimensional character and enhanced $p$–$d$ hybridization at Ni$_A$ promote greater electronic mobility, whereas Ni$_B$, with its more localized two-dimensional nature, exhibits a higher effective mass and reduced mobility.

\begin{table}[htbp]
\centering
\caption{Magnetic exchange parameters and anisotropy energy at the Ni sites as obtained from DFT calculations (GGA+$U$) and SpinW fitting of the experimental data.}
\label{tab:exchange_parameters}
\begin{tabular}{lccccc}
\hline\hline
Model & $J_1$ & $J_2$ & $J_3$ & $J_4$ & $A_{\zeta}$ (meV) \\
\hline
GGA+$U$ & $-4.16$ & $0.78$ & $-2.35$ & $18.65$ & -- \\
SpinW   & $-4.8(1)$ & $0.5(1)$ & $-1.54(3)$ & $27.9(8)$ & 
\makecell[l]{$0.7(2)$ (Ni$_A$) \\ $-2.13(2)$ (Ni$_B$)} \\
\hline\hline
\end{tabular}
\end{table}

\begin{figure}[t]
\centering
\includegraphics[width = 8 cm]{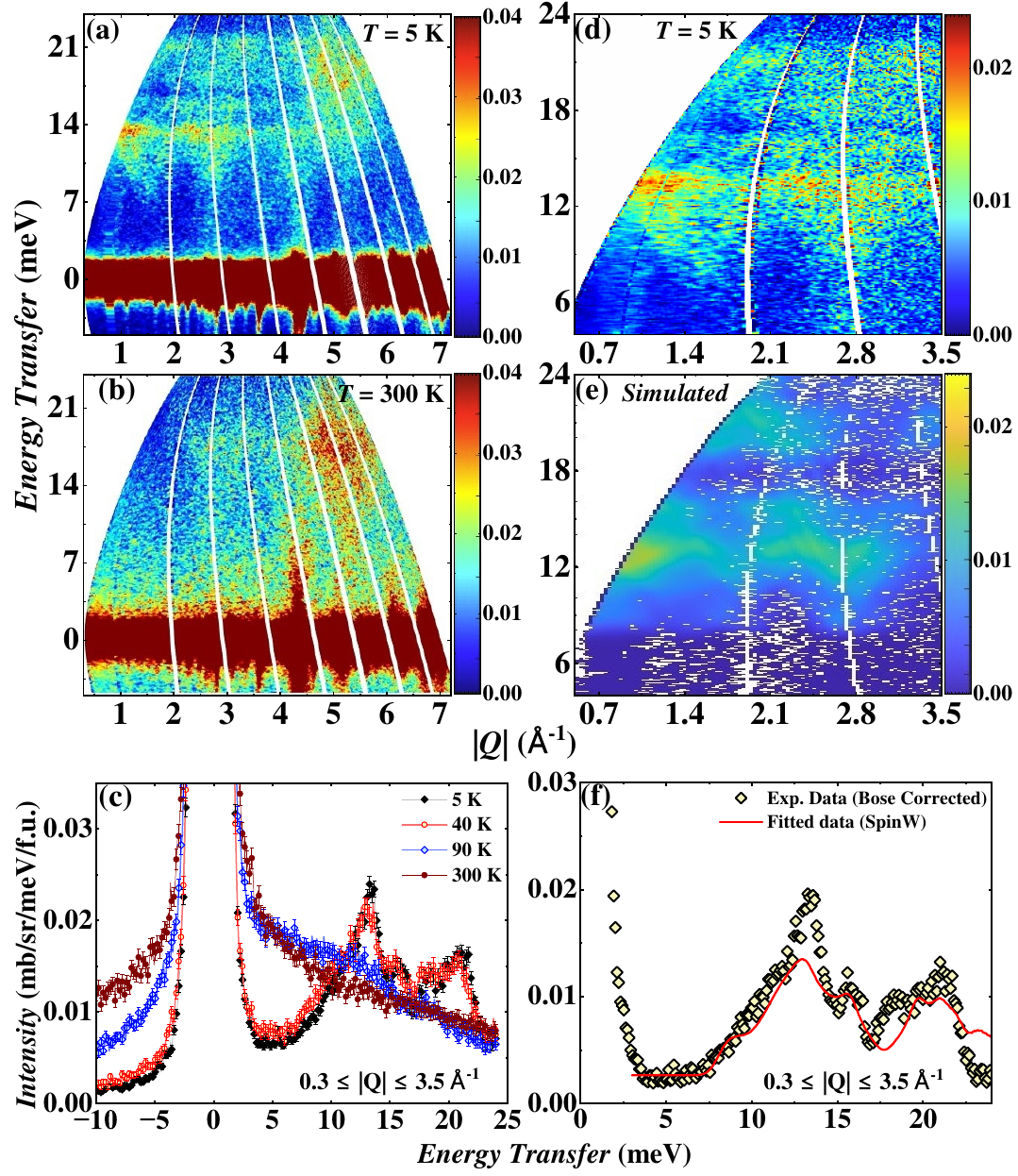}
\caption {(color online) 2D color plots of momentum transfer versus energy transfer obtained from INS experiments are shown in (a) and (b). (c) Scattering intensities integrated over the low-$|\mathbf{Q}|$ regime as a function of energy transfer at different temperatures. Phonon-corrected 2D color plots from (d) INS experiment at 5 K and (e) SpinW simulation. (f) Comparison between experimental and simulated scattering intensities as a function of energy transfer for $E_i =$ 30 meV.}  
\label{fig:INS}
\end{figure}    

INS measurements were carried out to investigate the nature of magnetic excitations in the system. Fig.~\ref{fig:INS}~(a) and (b) present the two-dimensional color plots at 5~K and 300~K, respectively, recorded with an incident energy of $E_i = 30$~meV. At 5~K, we observe two distinct bands of magnetic scattering centered in the energy ranges $\hbar\omega \approx 7$–$16$~meV and $16$–$22$~meV. These features vanish above the transition temperature $T_{\mathrm{FI}}$, confirming their magnetic nature. Sharp, dispersive features are observed in the ordered phase at momentum transfers of 1.4 and 2.8~\AA$^{-1}$, corresponding to the magnetic Bragg positions (111) and (222), respectively.  In particular, phonon modes appear at higher momentum transfers ($|\mathbf{Q}|$) in the same energy range, suggesting a coupling between lattice vibrations and spin degrees of freedom. A particularly striking feature is the presence of a significant gap ($\sim 7$~meV) in the spin excitation below $T_{\mathrm{FI}}$. The momentum-integrated energy cuts at various temperatures, shown in Fig.~\ref{fig:INS}~(c), reveal two prominent peaks centered at 13~meV and 20~meV, along with two shoulders at 15~meV and 18~meV. These excitations diminish on heating and disappear at 90~K. To model the low-temperature (5~K) Bose-corrected powder INS data with $E_i = 30$~meV, we employed the \texttt{SpinW v4.0} software package~\cite{spinw,SM} using its recently implemented powder-averaged fitting functionality based on linear spin wave theory. The spin Hamiltonian used is given by:

\begin{equation}
\mathcal{H} = \sum_{ij} J_{ij} \left( \vec{S}_i \cdot \vec{S}_j \right) + \sum_i A_{\zeta} \left(S_i^{\zeta} \right)^2
\end{equation}

The first term describes the isotropic Heisenberg exchange interactions, where $J_{ij}$ is the exchange strength between the $i$-th and $j$-th spins, and the second term accounts for single-ion anisotropy (SIA) along the direction $\hat{\zeta}$, parameterized by $A_{\zeta}$. Four exchange constants ($J_1$ to $J_4$) and the SIA parameters for the two crystallographically distinct Ni sites were optimized using a hybrid approach combining particle swarm optimization and the Levenberg–Marquardt algorithm, with constraints informed by DFT and point charge calculations. Instrumental resolution effects and kinematic constraints were incorporated into the simulations. Detailed methodology is provided in the Supplementary Material~\cite{SM}. The best-fit parameters show good agreement with the DFT predictions, as shown in Table~\ref{tab:exchange_parameters}, while the AFM interaction ($J_4$) between the two Ni sublattices turns out to be slightly stronger than the DFT estimates. For the Ni$_\mathrm{A}$ site, a weak easy-plane SIA of 0.7(2)~meV is observed in the $ac$-plane, while the Ni$_\mathrm{B}$ site exhibits a strong easy-axis SIA of -2.13(2)~meV along the $b$-axis. The order of magnitude different SIA arises in two different Ni sublattices from the different local environments at the Ni$_\mathrm{A}$ and Ni$_\mathrm{B}$ sites~\cite{Goddard}. This site-dependent anisotropy, which lacks a global axis or plane, competes with exchange interactions and influences the magnetic ground state. In three-dimensional and quasi-two-dimensional systems, alternating easy-axis anisotropy is known to induce spin canting and weak ferromagnetism, which are typically attributed to Dzyaloshinskii–Moriya interactions~\cite{Doll}. In contrast, alternating easy-plane anisotropy can stabilize collinear order along a pseudo-easy axis formed by the intersection of the local planes~\cite{Cliffe, Goddard1}. In NNO, the coexistence of easy-axis anisotropy at Ni$_\mathrm{B}$ and easy-plane anisotropy at Ni$_\mathrm{A}$, together with strong inter-sublattice AFM exchange ($J_4$), stabilizes a collinear FI ground state with spins aligned along the $b$-axis. SpinW simulations reveal that the large spin gap is primarily driven by the strong easy-axis anisotropy and the pronounced $J_4$ coupling. By systematically fitting the magnon spectra with varying SIA and exchange parameters ($J_1$–$J_4$), we reproduce both the excitation gap and observed sharp dispersions using a unique and well-constrained parameter set.


Our combined experimental and theoretical investigation reveals a distinct microscopic mechanism of FI in NNO, driven by subtle but significant differences in the local environments of two crystallographically inequivalent Ni$^{2+}$ sites. Despite identical valence and spin states, Ni$_\mathrm{A}$ and Ni$_\mathrm{B}$ sublattices exhibit contrasting octahedral distortions and magnetic dimensionality—Ni$_\mathrm{A}$ forms quasi-one-dimensional chains with pronounced distortions, while Ni$_\mathrm{B}$ forms nearly planar two-dimensional layers with minimal distortion. These differences lead to site-specific electronic structures: Ni$_\mathrm{A}$ exhibits a lower effective mass and enhanced $p$–$d$ hybridization, resulting in greater transfer of spin density to oxygen ligands and a reduced ordered moment, while Ni$_\mathrm{B}$ retains a near-full $S=$1 moment due to weaker hybridization and greater localization. Crucially, these local asymmetries give rise to alternating SIA-easy-plane at Ni$_\mathrm{A}$ and easy-axis at Ni$_\mathrm{B}$, which act as additional tuning parameters for the $S =$ 1 system. The easy-plane anisotropy at Ni$_\mathrm{A}$ confines spins to the $ac$-plane, reducing their projection along the orthogonal global magnetization direction, while the easy-axis anisotropy at Ni$_\mathrm{B}$ reinforces alignment along the $b$-axis, supporting the full spin-only ordered moment.  However, unlike conventional mechanisms based on mixed valency or multiple magnetic species, FI in NNO arises solely from structurally and electronically inequivalent sites of the same magnetic ion.

\par
J.S. acknowledges ANRF, India, for the Ramanujan Fellowship [Grant No. RJF/2019/000046(SQUID-1986-JS3632)] and POWER Grant [Grant No. SPG/2021/003136]. The authors thank Dr. Gavin Stenning and the Materials Characterization Laboratory at the ISIS Facility, Rutherford Appleton Laboratory, UK, for performing the heat capacity measurements. The authors thank Dr. Dmitry Khalyavin, ISIS Facility, Rutherford Appleton Laboratory, UK, for valuable scientific discussions.


\appendix

\setcounter{figure}{0}

\setcounter{table}{0}
\section{Methodology}
 A polycrystalline sample of Ni$_4$Nb$_2$O$_9$ (NNO) was synthesized using the standard solid-state reaction method~\cite{NBO1}. Preliminary powder x-ray diffraction performed at room temperature confirmed a single-phase material without detectable impurity peaks. Magnetic measurements were carried out using a Quantum Design superconducting quantum interference device (SQUID) magnetometer over the temperature range 2–300 K. Heat capacity measurements as a function of temperature were performed using a Quantum Design Physical Properties Measurement System under applied magnetic fields of 0 and 5 Tesla. X-ray absorption spectroscopy (XAS) measurements were carried out at the B-18 beamline, Diamond Light Source, UK, around the Ni-$K$ and Nb-$K$ edges in standard transmission geometry. Data were analyzed with freely available DMETER software packages~\cite{dmeter1,dmeter2}. 
 Powder neutron diffraction (PND) measurements were conducted at the ISIS Facility, Rutherford Appleton Laboratory, United Kingdom, using the WISH time-of-flight diffractometer~\cite{DOI}. Diffraction data were collected at temperatures ranging from 1.5 to 300 K. Approximately 7 g of powdered NNO was mounted in an 8-mm diameter vanadium can and cooled down to 1.5 K using a standard He cryostat. Rietveld refinements were performed on the diffraction data using the FullProf software package~\cite{Fullprof}. Analysis confirmed that NNO crystallizes with orthorhombic symmetry (space group $Pbcn$) and lattice parameters $a = 8.729(6)$\AA, $b = 5.078(1)$\AA, and $c = 14.304(3)$\AA at room temperature, consistent with previously reported data~\cite{NBO1, NBO2}. Inelastic neutron scattering (INS) experiments were performed on the time-of-flight spectrometer MARI at the ISIS Neutron and Muon Facility, Rutherford Appleton Laboratory, United Kingdom. Approximately 7 g of powdered NNO was placed inside a thin aluminium foil envelope (40 mm height and 140 mm length), rolled into a cylindrical shape, and inserted into an aluminium can (40 mm diameter, 0.1-mm wall thickness). The sample was cooled using a He-4 closed-cycle refrigerator. A straight Gd slit package was used in the Fermi chopper, rotating at frequencies of 200 Hz and 400 Hz to simultaneously record spectra at incident energies of 30 and 6.2 meV, and 120 and 25 meV, respectively, utilizing the repetition-rate multiplication method~\cite{nakamura,russina}. Data were collected at temperatures of 5, 40, 90, and 300 K. To obtain scattering intensities in units of cross-section (mb.sr$^{-1}$.meV$^{-1}$.fu$^{-1}$), vanadium spectra were measured under identical conditions. Data reduction was performed using MANTIDPLOT software~\cite{arnold}.
\par
The DFT calculations were conducted utilizing a plane-wave based basis set with a 500 eV cut-off within a pseudopotential framework employing the Perdew-Burke-Ernzerhof (PBE)~\cite{Perdew1996,pbe} exchange-correlation functional as implemented in the Vienna ${ab-initio}$ simulation package (VASP)~\cite{Kresse1993,Kresse1996}. To capture the missing electron-electron Coulomb correlations at the Ni-3$\textit{d}$ states, we employed an onsite Hubbard $U$ ($U_{eff}$=$U$-J$_H$)~\cite{Anisimov1993,Dudarev1998}. The SOC effect has been incorporated in the calculations through relativistic corrections to the original Hamiltonian ~\cite{Hobbs2000,Steiner2016}. We used $10\times8\times10$ k-mesh in the Brillouin zone (BZ) for the self-consistent calculations. The structural optimization was performed by relaxing the atomic positions towards equilibrium until the Hellmann-Feynman force becomes less than 0.001 eV/$\AA$, keeping the lattice parameters fixed at their experimental values.
\begin{figure*}[t]
\centering
\includegraphics[width = 15 cm]{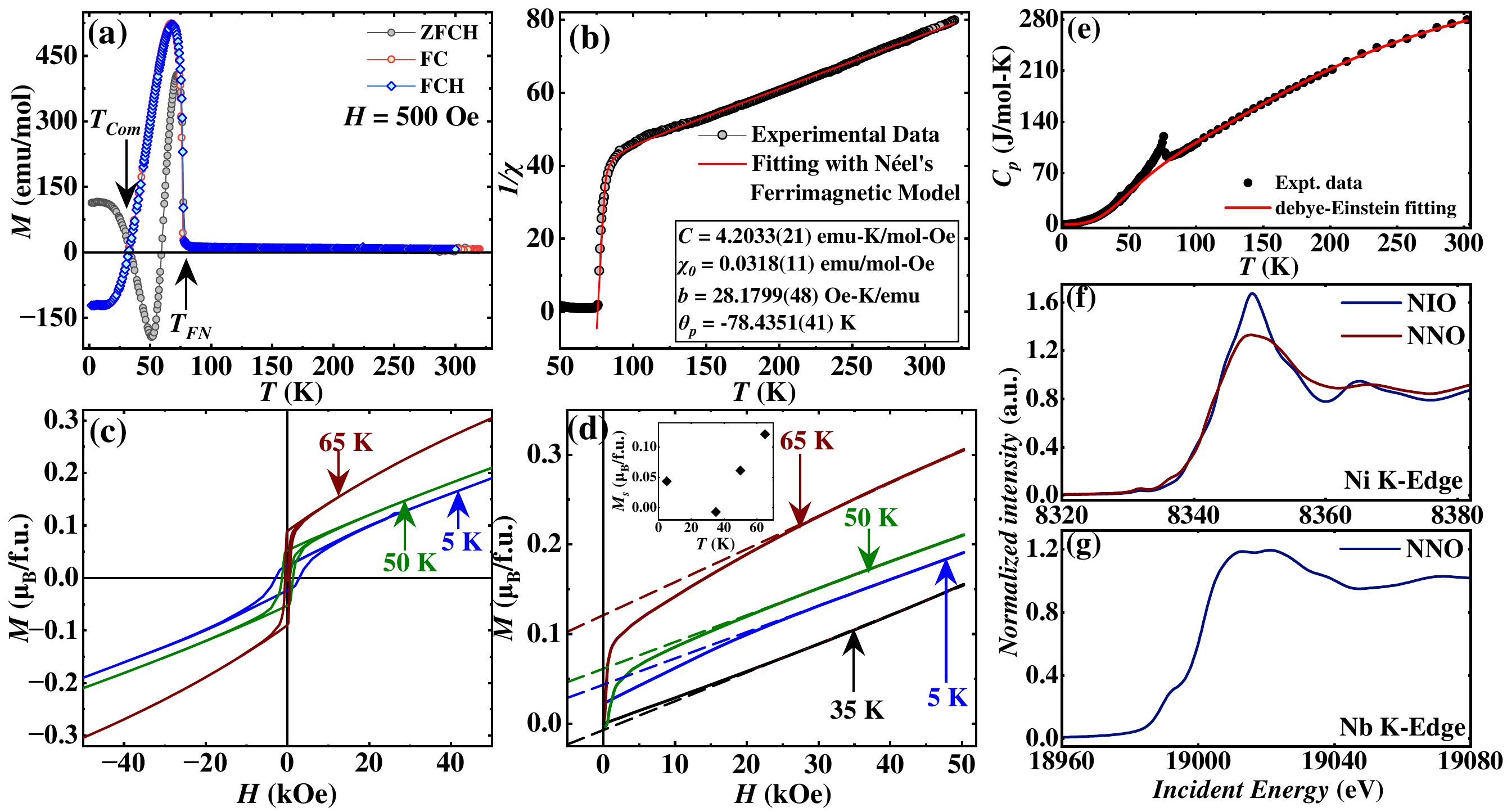}
\caption {(color online) (a) Temperature dependence of ZFCH, FC, and FCH magnetization data. (b) Inverse susceptibility (1/$\chi$) as a function of temperature, along with a fit using the N\'{e}el ferrimagnetic model. (c) Isothermal magnetization ($M$) versus magnetic field ($H$) data at 5, 50, and 65 K. (d) The first quadrant of the $M$ vs. $H$ data with a linear fit in the high magnetic field region at 5, 35, 50, and 65 K, with the inset displaying the temperature dependence of the $M_s$.(e) Shows heat capacity of NNO as a function of temperature. The scattered points are experimental data, while the solid line is the lattice heat capacity.Ni-$K$ edge and Nb-$K$ XANES data are shown in (f) and (g) respectively.}  
\label{fig:magnetic}
\end{figure*}    
\section{Bulk Magnetic Properties}

Fig.~\ref{fig:magnetic}(a) presents the magnetization ($M$) versus temperature ($T$) data collected under zero-field-cooled-heating (ZFCH), field-cooled (FC), and field-cooled-heating (FCH) protocols with an applied magnetic field of $H =$ 500 Oe. The $M(T)$ curve exhibits a sharp increase at $T_{FN} \approx$ 78 K, signalling the onset of ferrimagnetic ordering, with a compensation temperature $T_{com} \approx$ 33 K. In Fig.~\ref{fig:magnetic} (b), the temperature dependence of the paramagnetic susceptibility $\chi$ for $T > T_{FN}$ is analyzed using the N\'{e}el model for ferrimagnetism which yields the temperature dependence of $\chi$ versus $T$ as:

 \begin{equation}
     \frac{1}{\chi} = \frac{T + \frac{C}{\chi_0}}{C} - \frac{b}{T - \theta}
 \end{equation}
Here, $C$ is the Curie constant, $\chi_{0}$ is the temperature-independent component, and $b$ incorporates the Weiss molecular field information for intra- and inter-sublattice exchange fields. This expression forms a hyperbola intersecting the temperature axis at the paramagnetic Curie point ($\theta_p$). The second term diminishes at high temperatures, reducing the equation to the Curie-Weiss law. From this fitting, we determined the effective paramagnetic moment as 2.89 $\mu_B$/Ni2+, which closely matches the spin-only moment for $S =$ 1 (2.828 $\mu_B$ ). Isothermal magnetization ($M$ versus $H$) measurements were conducted at temperatures ranging from 5 K to 100 K. Representative data at $T =$ 5 K, 50 K, and 65 K are shown in the main panel of Fig.~\ref{fig:magnetic}(c), with the 100 K data displayed in the inset. The main panel of Fig.~\ref{fig:magnetic}(d) illustrates the first quadrant of the $M(H)$ curves at various temperatures below $T_{FI}$. The high-field data are fitted using the empirical relation $M(H) = \chi_{AFM}H + M_S$, where $\chi_{AFM}$ is the linear term corresponding to the antiferromagnetic component, and $M_S$ is the spontaneous magnetization arising from the inequivalent magnetic moments between the two Ni sublattices. Near $T_{com}$ (at 35 K), the $M$ versus $H$ curve is almost linear, indicating the vanishing $M_S$, as depicted in the inset of $M_S$ versus $T$ plot. The $M_S$ reaches its maximum at 65 K, indicating the larger magnetic moment difference between the two Ni sublattices just below $T_{FI}$.
\begin{figure*}[t]
\centering
\includegraphics[width = 15 cm]{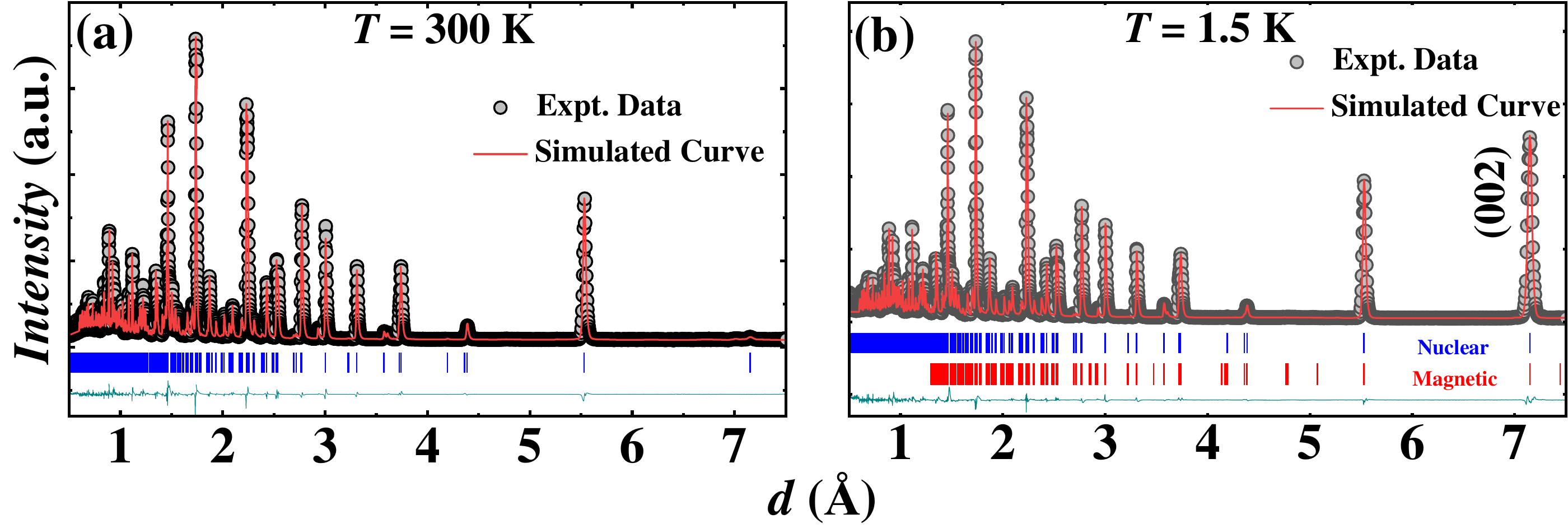}
\caption {(color online) PND patterns along with corresponding Rietveld refinements at 300K and 1.5K are shown in (a) and (b), respectively. Blue and red vertical ticks denote nuclear and magnetic Bragg peak positions, respectively. The solid green line represents the difference between the experimental data and the simulated fit.}  
\label{fig:PND}
\end{figure*}    
\begin{figure*}[t]
\centering
\includegraphics[width = 16 cm]{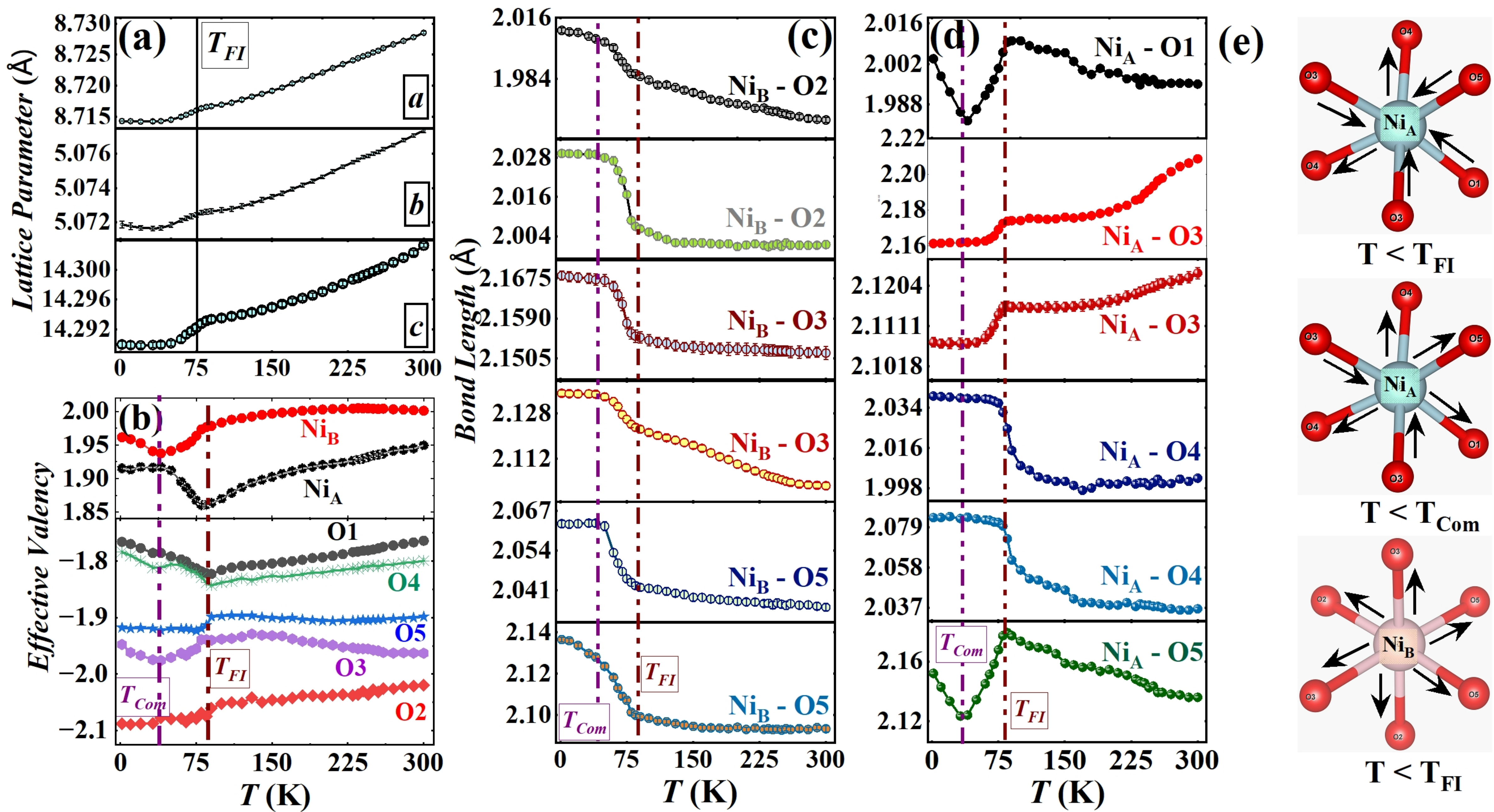}
\caption {(color online) (a) Temperature dependence of the lattice parameters. (b) Effective valencies of Ni and O as a function of temperature, obtained from bond valence sum (BVS) calculations. (c,d) Temperature evolution of the Ni$_B$–O and Ni$_A$–O bond lengths, respectively. (e) Perspective views of Ni$_A$O$_6$ and Ni$_B$O$_6$ octahedra, illustrating bond-length variations with temperature.}  
\label{fig:BVS}
\end{figure*}    

\section{Heat Capacity}
Fig.~\ref{fig:magnetic}(e) presents the temperature dependence of the heat capacity ($C_P$) of NNO measured between 2 and 300~K. The total heat capacity can be expressed as the sum of lattice ($C_{\mathrm{lat}}$) and magnetic ($C_{\mathrm{mag}}$) contributions: $C_P = C_{\mathrm{lat}} + C_{\mathrm{mag}}$. Given the insulating nature of the sample, electronic contributions were neglected. In the absence of a suitable nonmagnetic isostructural reference compound, $C_{\mathrm{lat}}$ was estimated by fitting the high-temperature ($T > 100$~K) data. A simple Debye model was insufficient to describe the high-temperature behavior; thus, a combined Debye--Einstein model~\cite{martin-hc} was employed, defined as:
\begin{equation}
C_P(T) = \alpha\,C_D(T,\theta_D) + (1 - \alpha)\,C_E(T,\theta_E)
\end{equation}
where $C_D(T,\theta_D)$ and $C_E(T,\theta_E)$ represent the Debye and Einstein contributions, respectively, and $\theta_D$ and $\theta_E$ are the corresponding characteristic temperatures. The best fit, obtained in the temperature range 100-300~K, provided $\alpha = 0.43$, $\theta_D = 300$~K, and $\theta_E = 782$~K. Below approximately 80~K, deviations from the fitted model were observed, indicating additional low-temperature contributions. The fitted curve was extrapolated to lower temperatures to represent $C_{\mathrm{lat}}$, from which the magnetic contribution was subsequently calculated as $C_{\mathrm{mag}} = C_P - C_{\mathrm{lat}}$.

\section{ X-ray absorption near-edge structure (XANES)}
The charge state of the magnetic cations plays a crucial role in determining the magnetic interactions within the system. To estimate the valence states of Ni and Nb, we analyzed the XANES of NNO. The Ni-$K$ edge XANES spectra, shown in Fig.~\ref{fig:magnetic}(f), confirm that Ni is in the 2+ oxidation state, as it closely follows the rising line features and the pre-edge position of the NiO standard. In contrast, the Nb XANES spectra (see Fig.~\ref{fig:magnetic}(g)) exhibit a broad hump-like feature around 18,990 eV, similar to that observed in Nb$_2$O$_5$~\cite{Nb-XANES}, indicating that Nb is in the 5+ oxidation state in the system. This also affirms the stoichiometry of the present compound. 

\begin{table*}[htbp]
\centering
\caption{Model SpinW Optimiser: Fitting parameters using PSO and LM optimizers in SpinW. Exchange parameters $J_1$–$J_4$, anisotropies $A$ and $B$ are in meV. Cost quantifies the fit quality.}
\label{tab:spinw_model_fits}

\scriptsize
\begin{tabular}{lccccccccccc}
\hline\hline
 & $J_1$ & $J_2$ & $J_3$ & $J_4$ & $A$ & $B$ & Cost & nRand & Hermit & diff\_step \\
\hline
Min           & $-7$      & $-2$     & $-3$     & $10$      & $-1$     & $-3$     &         &       &        &         \\
Max           & $-3$      & $2$      & $-1$     & $30$      & $1$      & $-0.1$   &         &       &        &         \\
PSO (Global)  & $-4.66$   & $0.62$   & $-1.54$  & $28.3$    & $0.78$   & $-2.17$  & $0.7195$ & $7e2$  & false  & --      \\
LM4 (Local)   & $-4.78(5)$& $0.52(2)$& $-1.52(4)$& $28.4(1)$ & $0.7(1)$ & $-2.12(2)$ & $0.7176$ & $2e3$  & false  & $1e^{-2}$ \\
LM4 (Local)   & $-4.78(9)$& $0.48(5)$& $-1.54(4)$& $27.9(6)$ & $0.7(1)$ & $-2.13(3)$ & $0.7179$ & $1e4$  & false  & $5e^{-3}$ \\
LM4 (Local)   & $-4.8(1)$ & $0.5(1)$ & $-1.54(3)$& $27.9(8)$ & $0.7(2)$ & $-2.13(2)$ & $0.7189$ & $1e4$  & true   & $5e^{-3}$ \\
\hline\hline
\end{tabular}
\end{table*}
\section{Powder Neutron Diffraction (PND)}

The full PND patterns of NNO at 300 K and 1.5 K, obtained from the low-scattering angle detector banks, are shown in Figs.~\ref{fig:PND}(a) and (b), respectively. The difference between the PND patterns at 1.5 K and 300 K (below and above $T_{FI}$, respectively) reveals no additional resolution-limited Bragg peaks. The PND patterns recorded at different temperatures show increasing intensity on top of existing Bragg peaks below $T_{FI} =$ 76 K. The strongest increase in intensity is observed for the (002) peak at a large $d$-spacing ($d =$ 5.48 \AA), indicating its magnetic origin. All the magnetic reflections can be indexed using the $k =$ (0 0 0) propagation vector, consistent with the previous PND study ~\cite{NBO3}. Magnetic symmetry analysis for determining the magnetic structure results in space group $Pbcn$, with the $k =$ (0 0 0) propagation vector and the magnetic Wyckoff site for Ni1 and Ni2 sites (8$d$), in eight one-dimensional irreducible representations: 3($\Gamma1 + \Gamma2 + \Gamma3 + \Gamma4 + \Gamma5 + \Gamma6 + \Gamma7 + \Gamma8$). Among these, only $\Gamma5$ indicates the ferromagnetic (FM) configuration within the sublattice of Ni$_A$ and Ni$_B$.

\section{SpinW}
The Bose-corrected low-temperature (5~K) INS powder data measured on MARI with incident energy $E_i = 30$~meV were fitted using recently implemented functionality in SpinW v4.0; this is the first publication to employ this capability. SpinW simulates the INS cross-section of the spin-wave spectrum using linear spin-wave theory (LSWT), as described in Ref.~\cite{spinw}. The powder-averaged spin-wave spectrum is calculated by numerical integration at random points distributed over a spherical surface of radius $|Q|$ around the origin in reciprocal space for each considered $|Q|$. Numerical integration was performed using a lattice of $n$ sampling points based on a pair of successive Fibonacci numbers, as detailed in Ref.~\cite{fibonacci}. The kinematic range of the instrument for a given $E_i$ was accounted for by specifying a minimum scattering angle measured (the maximum $|Q|$ of the fitted data is well below the high-angle kinematic cutoff). The energy and momentum ($Q$) resolutions were assumed to be separable. The energy resolution was treated as a Gaussian function with a full width at half maximum (FWHM) calculated analytically using the MARI instrument model within the PyChop package~\cite{pychop}, and was used to weight the structure factors at each eigenvalue contributing to a given energy bin at a specific $|Q|$. The momentum dependence of the resolution was modelled by applying a Gaussian broadening in $|Q|$, with the corresponding FWHM determined from an analytical model considering the angular widths of the instrument components. The fitting was performed using an unweighted least-squares cost function, evaluating residuals within the range $|Q| = 0.5$--$2.5$~\AA$^{-1}$ to avoid strong phonon scattering at higher $|Q|$, and an energy-transfer range of $4$--$24$~meV to exclude contamination by (quasi-)elastic scattering. The exchange parameters ($J_1$--$J_4$) and the single-ion anisotropies ($A$ and $B$) at the two Ni sites—corresponding to easy-axis/easy-plane anisotropies aligned along or normal to the $b$-axis—were optimized, along with a multiplicative scale factor and a constant background. Global optimization was first conducted using the particle-swarm optimizer (PSO) implemented in SpinW, employing a relatively small number of sampling points for the powder average ($n = 7\times10^2$). Parameter ranges were constrained based on coarse grid searches centred around exchange estimates obtained from DFT calculations and single-ion anisotropies derived from point-charge calculations. Subsequently, the minimum identified by PSO was refined via a Levenberg-Marquardt (LM) fitting procedure using a larger number of sampling points ($n = 1\times10^4$). The LM fit converged to the same minimum identified by PSO. Parameter uncertainties were calculated from the diagonal elements of the covariance matrix derived from the Hessian estimated during the LM minimization. The optimized parameters are summarized in Table~\ref{tab:spinw_model_fits}.

 \end{document}